\documentclass{ieeeclass}
\usepackage{xcolor}
\usepackage{tabu}
\usepackage{graphicx}

\ieeetitle{Mobile Edge Computing and Artificial Intelligence:\\ A Mutually-Beneficial Relationship}

\author{
	Ahmed A. Al-habob and  Octavia A. Dobre
	 
	 Faculty of Engineering and Applied Science, Memorial University,  Canada
	 
	 Email: \{\textit{aaaalhabob, odobre}\}@mun.ca\authorcr

}

\begin{document} 

\maketitle 

\ieeeabstract{This article provides an overview of mobile edge computing (MEC)  and
	artificial intelligence (AI) and discusses the mutually-beneficial relationship between them. 
	AI provides revolutionary    solutions      
	in nearly every important aspect of the
	MEC offloading process, such as resource management and scheduling. On the other hand, MEC servers  are utilized to avail a distributed and parallelized learning framework,  namely mobile edge learning.  
}

\section{Introduction}
Computing is increasingly ubiquitous
 in all aspects of our life, which adds additional functions to  phones,  tablets, and wearable devices.\ Computing requirements for mobile applications are increasing rapidly to meet the need for computational-extensive applications of mobile users, such as virtual reality and real-time online gaming.\ Mobile edge computing (MEC) is a promising solution to   avail adequate computing and storage
 capabilities in close
 proximity to mobile users [\ref{R1}].\ The main target of previous generations wireless networks   was to provide   adequate wireless speeds to make the transition
 from voice-centric to multimedia-centric viable.\ The mission of 5G networks
 is quite  different and much more complex, namely to uphold    communications, computing,
 control and content delivery.
 
In 5G wireless networks, ultra-dense edge devices, such as wireless access points and  
small-cell base stations are deployed, each
 providing  computation and storage capabilities
 at the network edge.\ Moreover, unmanned aerial vehicles  have attracted significant attention to provide aerial-ground cooperative MEC frameworks 
due to  their  agile management, flexible deployment,
and low cost [\ref{R7}].

Considering such diverse resource availability and  the massive amount of devices and data  generated by the computationally intensive applications, there is a   need for powerful tools
and techniques capable of  allocating both communications and computing resources to  users.\ Artificial intelligence (AI) solutions,   in particular   deep learning  networks, represent a fit to address the hurdle and empower intelligent resource management  for efficient MEC in real-time and dynamic scenarios.  

Recently, the concept of  mobile edge learning (MEL) has been defined, in which MEC interplays with AI in the sense that the learning model, parameters, and  
data are distributed across multiple edge servers, and
an AI model is trained from  distributed data [\ref{R3}].\ Such a distributed  learning  model is known as federated learning, in which a node   plays the role of the orchestrator that aggregates   locally derived parameters   
and returns globally updated parameters to the servers (learners) [\ref{R5}].\ Such a  mutually benefiting interaction between MEC and AI  paves the way for an intelligent-pipe, in which the communication network becomes intelligent and self-driven.

\section{Mobile Edge Computing}
 The first real-world MEC platform---introduced by Nokia Networks  and Intel---aims to support  the base station with computation capability to providing  intelligent services  and  collecting real-time network data such as subscriber locations and cell congestion [\ref{R1}].\ Saguna and Intel introduced a fully virtualized   software-based MEC platform,
   which can provide an open
environment for running third-party MEC applications [\ref{R8}].\ The European Telecommunications Standards Institute formed 
   proofs-of-concepts  to demonstrate the viability of MEC implementations as a key technology of the 5G  era [\ref{R9}].    

To be a viable and competitive option, the offloading process in a MEC framework should be implemented with    low energy consumption,   low offloading error,      and low latency.  
A large body of literature   has studied different offloading  policies to address these requirements,  such as: \textit{binary} offloading, in which the mobile device
determines whether a   task should be offloaded to the MEC servers or  computed locally; and \textit{partial} offloading, in which a portion of the computation is performed locally at the mobile device   and the other portion is  offloaded to the MEC servers. More evolved policies have also been studied,  in which  the  task of a mobile device   is partially processed at cloud servers and    partially offloaded to the  edge servers for computing.\ Both parallel and sequential offloading schemes have been addressed in the literature [\ref{R6666}].

The massive amount of edge devices and great variety of applications and services   make  resource management in the MEC framework a complex process.
Even in the simple scenarios of offloading  
single user's tasks or offloading to a single server, the offloading decision    problem is  NP-hard  or NP-complete  [\ref{R6}].
Moreover,  the following practical issues need to be addressed in the computational task offloading process:
\begin{itemize}
	\item \textit{Prioritized tasks}: In practical scenarios, some tasks (or users)  require higher priority to access the computation resources;
	\item \textit{Dependency} among  tasks: In many applications, the computation of a task depends on the computed results of other tasks.   Inter-tasks   dependency  cannot be ignored and  has a significant  effect on  the offloading    and	computation procedure;
	\item \textit{Repeated tasks}: More than one user may offload the same task, and such a duplicated effort reduces the effectiveness of the offloading process.   Collaborative offloading schemes could reduce the offloading cost and improve the offloading process.
\end{itemize}
With such requirements and potentially large    number of variables,  
MEC offloading  ends up being a high complexity and dimensionality  problem, which requires powerful and real-time resource  management algorithms with 
  ability to respond to changes in the offloading environment.
\section{Artificial Intelligence Approaches for MEC}
AI can be viewed as a
collection of  algorithms that empower intelligent
machines to improve their performance on  descriptive, predictive, and prescriptive models. Conventional  AI approaches may be good enough for handling moderate-dimensional situations; however, the  MEC resource  management problem requires dynamic solutions and  ability of dealing with
high dimensionality   scenarios efficiently.\ For example, reinforcement learning  has shown promising
results  for resource management, such as
abstract computing or memory resources [\ref{R10}] and  production scheduling and control [\ref{R11}].\ First, the resource management and scheduling  problem is modeled as a
Markov decision process.\ Then, a reinforcement technique is used in which an agent   finds optimal
actions in order to optimize the performance by maximizing the
reward of each action. The agent trains a  deep neural network---an artificial neural network   with multiple layers between the input and output layers---by  performing an iterative state transition process and characterizing each state with main features.\ The iterations continue until the agent decides to   use the best   action found so far.

\begin{figure}[h]
	\centering
	\includegraphics[width=\linewidth]{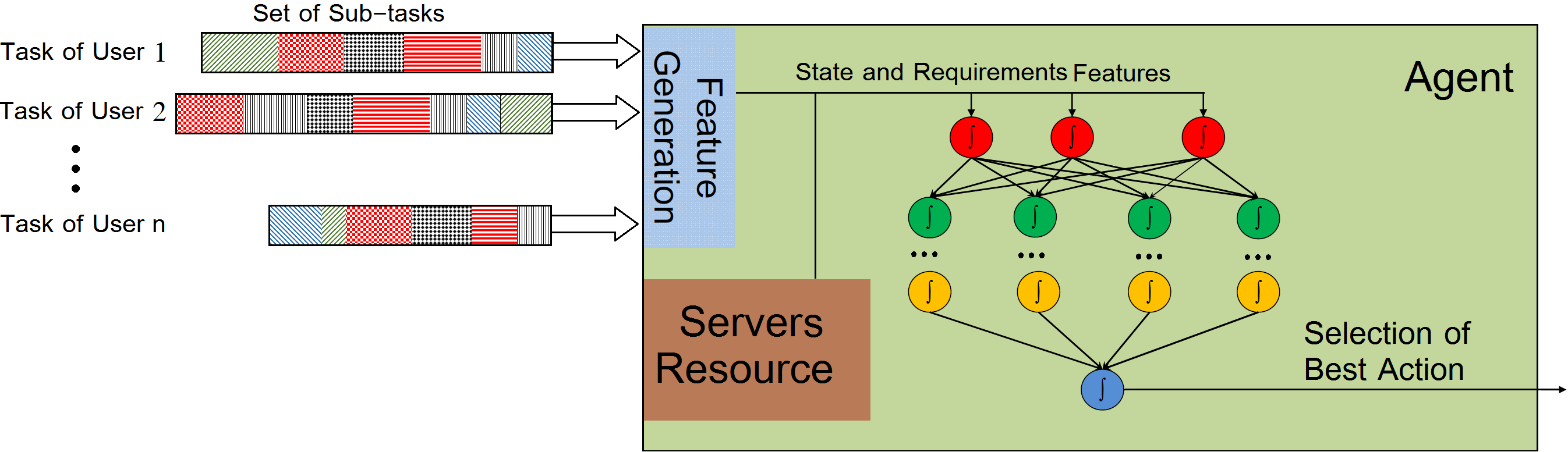}
	\caption{An agent in which the 
		inputs of the deep learning network are   features describing the servers' current status and  users' requirements.}
	\label{figfig1}
\end{figure}

Let us consider the scenario in which     the best offloading decision needs to be found, which associates 
each task generated by  mobile users with one of the available MEC servers, with the goal of minimizing
the offloading error, latency, energy consumption, or a combination of these objectives. In this case,    the state of a server can be described by its resources such as computing speed and   communication channel quality.\ User  requirements   include    tasks payload size and the   required  
processing cycles; dependency and/or priority among the tasks/sub-tasks also needs to be taken into account.\ Then, an agent iteratively associates the users' tasks  with  servers  and records the resulting reward, which is the reduction of the objective.\ The features describing
the status of the server and the   requirements of users are fed to  learn  the  deep neural network.\ The generalization  capabilities of the deep neural network yield  general scheduling policies  which are not just tuned to the states encountered
during training, but  are adaptive   to be applied to unknown states,
too.
Figure \ref{figfig1} illustrates the basic components of an agent. 

 For a multi-agent scenario, multiple interacting intelligent agents   collaborate  to  reinforce the learning process that results in a lower objective.
The
critical part in designing a reinforcement learning algorithm is to
define a reward function that captures the offloading requirements and governs the algorithm  performance.

\section{From Edge Computing to Edge Learning}
\begin{figure}[h]
 	\centering
 	\includegraphics[width=1\linewidth]{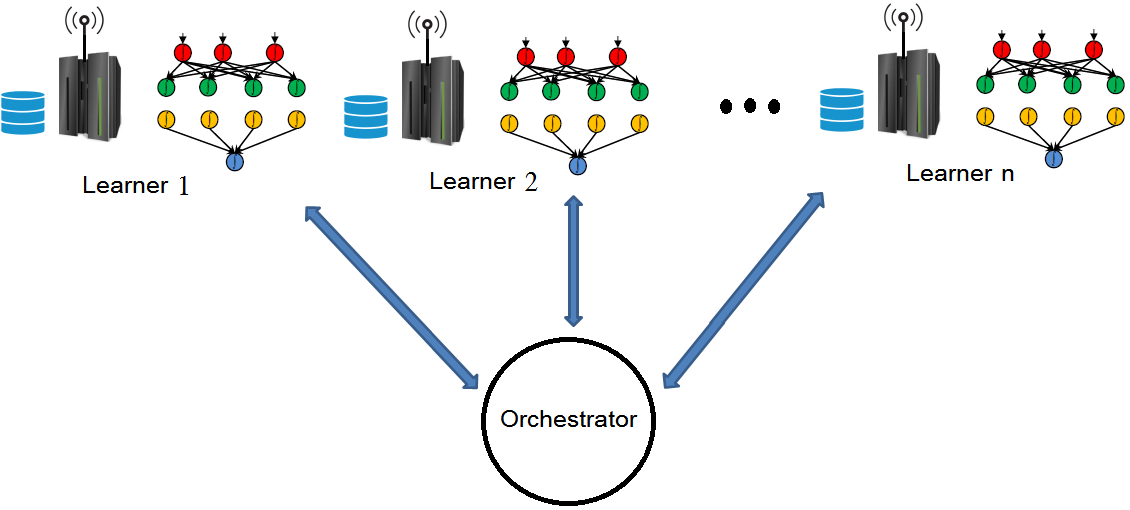}
 	\caption{MEL model.}
 	\label{figfig2}
 \end{figure}

In many realistic applications,  an AI algorithm 
is a computationally-expensive  task and requires    large-scale training samples.\ The  convergence rate of the learning process can be improved  using the federated learning technique in which     the learning process and the training samples are parallelized between processing nodes (a.k.a., learners).\ The massive deployment of  servers  at the network edge motivates the concept of MEL, in which  the computationally-expensive AI algorithm is distributed  and executed on
  the edge servers [\ref{R5}].\ In the MEL  framework, each server (learner) performs training iterations to train  its local learning model using local training samples.\ Once each learner finishes its iterations, it forwards
  the resulting features to the orchestrator.\ The orchestrator---which is a
  logical component that can run on a remote cloud, network
  element, or edge node---then aggregates the
  local features   and updates all learners. 
  MEL enables the AI algorithm    gain diversity from the vast range of  data located at different servers.\ It is worth mentioning that the learners and orchestrator exchange  the extracted features, whereas the raw training data  remain on the learners.\ The extracted features are  small in size in comparison with the raw  training data; consequently,  transmitting the extracted features     reduces the burden on the communication channels. Figure \ref{figfig2} illustrates the general  MEL model.     

It is envisioned that beyond 5G wireless networks   will be required to support ubiquitous AI services from the core to the end devices of the network.\ AI will play a critical role
in designing and optimizing beyond 5G wireless networks architectures, protocols, and
operations.\ Meanwhile,  the design of the beyond 5G wireless networks architecture will follow an \textquotedblleft AI native\textquotedblright\ approach, where intelligentization will allow the network to
be smart, agile, and able to learn and adapt itself according
to the changing network dynamics [\ref{R55}].

\newpage
\section{Conclusion}
This article has briefly presented the basic concepts,  benefits, and integration of MEC and AI. 
The potential of incorporating 
extended AI functionalities in MEC has been discussed. The idea of MEL has been introduced, in which  MEC servers  are utilized to distribute and parallelize  the resource-intensive  learning process of an AI model.\ It is envisioned that AI at the edge will be  an integral part of beyond 5G wireless networks.

\section*{References}
\begin{enumerate}
\item \label{R1} Intel and Nokia Siemens Networks,\ \textquotedblleft Increasing mobile operators' value proposition with edge computing,\textquotedblright \textit{Technical brief}, 2013.
\item \label{R7} N. Cheng  et al.,\ \textquotedblleft Air-ground integrated mobile edge networks: Architecture, challenges, and opportunities,\textquotedblright \textit{IEEE Commun. Mag.}, vol.  56, no. 8, pp. 26-32, Aug. 2018.
\item \label{R3} S. Wang et al.,\ \textquotedblleft Adaptive federated learning in resource constrained edge computing systems,\textquotedblright \textit{IEEE J. Sel. Areas Commun.}, vol. 37, no. 6, pp. 1205-1221, Jun. 2019.
\item \label{R5} U. Mohammad    and S. Sorour,\ \textquotedblleft Adaptive task allocation for mobile edge learning,\textquotedblright \textit{arXiv preprint arXiv:1811.03748},   Nov. 2018.	
\item \label{R8} Saguna and Intel,\ \textquotedblleft Using mobile edge computing to
improve mobile network performance and profitability,\textquotedblright
\textit{White paper}, 2016.
\item \label{R9} Y. C. Hu et al.,\ \textquotedblleft Mobile edge computing - A key technology towards 5G,\textquotedblright \textit{ETSI White paper}, vol. 11, 2015. 
\item \label{R6666} A. Al-habob et al.,\ \textquotedblleft Collision-free sequential task offloading for mobile edge computing,\textquotedblright \textit{IEEE Commun. Lett.},  Early Access, Oct. 2019.
\item \label{R6} T. K. Rodrigues et al.,\ \textquotedblleft Machine learning meets computation and communication control in evolving edge and cloud: challenges and future perspective,\textquotedblright   \textit{IEEE Commun. Surveys Tut.},  Early Access, Sep. 2019.
\item \label{R10}  A. I. Orhean et al.,\ \textquotedblleft New scheduling approach using reinforcement learning for heterogeneous distributed systems,\textquotedblright   \textit{J.   Parallel and Distrib. Comput.}, vol. 117,   pp. 292-302, Jul. 2018.
\item \label{R11}  Y. R. Shiue et al.,\ \textquotedblleft Real-time scheduling for a smart factory using a reinforcement learning approach,\textquotedblright \textit{Computers \& Industrial Engineering}, vol. 125   pp. 604-614, Nov. 2018.	
\item \label{R55}  K. B. Letaief et al.,\ \textquotedblleft The roadmap to 6G--AI empowered wireless networks,\textquotedblright \textit{arXiv preprint arXiv:1904.11686},   Apr. 2019.	
\end{enumerate}

\newpage
\ieeeauthorpicture{Ahmed}
\ieeeauthorcv{\textbf{Ahmed A. Al-habob} (S'15) received the BSc degree in telecommunications and computer engineering from Taiz Univ-\\ersity,\ Yemen,\ in 2009.\ He received the MSc degree in telecommunications from the Electrical Engineering Department, King Fahd University of Petroleum and Minerals (KFUPM), Saudi Arabia, in 2016.\ He is currently a PhD student at the Faculty of Engineering and Applied Science, Memorial
	University, St. John’s, NL, Canada.\ His research interests include wireless
	communications and networking.}
\vspace{.5cm}
\ieeeauthorpicture{OctaviaDobre}
\ieeeauthorcv{\textbf{Octavia A. Dobre}  (M'05-SM'07) is a Professor and Research Chair at Memorial University, Canada. She was a Visiting Professor at Massachusetts Institute of Technology, as well as a Royal Society and a Fulbright Scholar. Her research interests include technologies for 5G and beyond, as well as optical and underwater communications. She published over 250 referred papers in these areas. Dr. Dobre serves as the Editor-in-Chief (EiC) of the IEEE Open Journal of the Communications Society. She was the EiC of the IEEE Communications Letters, a senior editor and an editor with prestigious journals, as well as General Chair and Technical Co-Chair of flagship conferences in her area of expertise. She is a Distinguished Lecturer of the IEEE Communications Society and a fellow of the Engineering Institute of Canada.}

\end{document}